# EVOLUTION FEATURES OF GIANT SOURCES WITH LINEAR AND BREAK STEEP RADIO SPECTRA

A.P. Miroshnichenko

Institute of Radio Astronomy, NAS of Ukraine,

Kharkiv, Ukraine, *a.p.miroshnichenko@gmail.com*

**ABSTRACT.** In the framework of the cosmological model ΛCDM the features of properties of giant radio sources with steep low-frequency spectra of linear (S) and break (C+) types are examined. Our estimates of characteristic age of galaxies and quasars with steep spectrum from the UTR-2 catalogue have values $10^7$-$10^8$ years. We consider that steep radio spectra may be formed at the transient injection by synchrotron losses of relativistic electrons. Earlier we have determined two evolution branches in relation of the characteristic age and ratio of emission of torus and accretion disk crown for sources with steep spectra S and C+. To further examination of evolution features of giant sources with steep spectra we consider ratio of emission of radio lobes and accretion disk (and disk crown) versus velocity of jet propagation and characteristic age of these objects. Analysis of obtained relations testifies to periodical activity of giant lowfrequency steep spectrum radio sources.

**Keywords:** Galaxies – Quasars – Radio sources: giants

## 1. Introduction

Non-thermal spectra of extragalactic sources have a number of features at the low-frequency band. Near 30 per cent of objects from the UTR-2 catalogue of extragalactic sources have steep radio spectra (the value of spectral index is greater 1) (Braude et al, 1981, 2003). To consecutive study of identified sources with low-frequency steep radio spectra we use corresponding data (http://ned.ipac.edu) at high–frequency range – centimetre, infrared, optical, X-ray bands. This gives the advantage in obtaining of many astrophysical characteristics of objects and in determination of their evolution effects. We have identified 130 galaxies and 91 quasars with steep low-frequency spectrum at selection criteria (low-frequency spectral index exceeds 1, and flux density is more than 10 Jy at the frequency 25 MHz) in two fields of UTR-2 catalogue. So, the compiled sample includes 78 galaxies with linear steep spectrum (S) and 52 galaxies with break steep spectrum (C+), 55 quasars with linear steep spectrum (S) and 36 quasars with break steep spectrum (C+) (Miroshnichenko, 2014). We use for calculations of physical parameters of considered sources the ΛCDM – Universe model with cosmological parameters $\Omega_m = 0.27$, $\Omega_\Lambda = 0.73$, $H_0 = 71$ km/s·Mpc.

The goal of this work is further examination of relations of emission contribution of structure components of giant steep-spectrum sources versus other physical characteristics.

## 2. Evidences for long evolution of sources with steep low-frequency spectra

When study of radio sources with steep low-frequency spectra from the UTR-2 catalogue we have determined that all these objects have giant structure (~ Mpc) (Miroshnichenko, 2012). Our estimates of characteristic age of galaxies and quasars with steep spectrum from the UTR-2 catalogue have values $10^7 - 10^8$ years (Miroshnichenko, 2013). At these great time scales the steep radio spectra of sources may be formed at the transient injection of relativistic electrons by their long synchrotron losses (Kardashev, 1962).

From the relation of initial and present spectral indices (Gorbatsky, 1986) we derive estimation of time ($t_{STI}$) necessary for observed steepness of radio spectrum of source at the case of transient injection of relativistic electrons:

$$t_{STI} = \frac{\left(\frac{\nu_2}{\nu_1}\right)^{(\alpha-\alpha_0)/(2\alpha_0-1)} - 1}{\mu B_x^2 \left[\left(\frac{\nu_2}{\nu_1}\right)^{(\alpha-\alpha_0)/(2\alpha_0-1)} E_2 - E_1\right]} \quad (1)$$

where $\alpha_0$ - is initial spectral index,

$\alpha$ - is present spectral index,

$\nu_1, \nu_2$ – are limit frequencies of the frequency interval,

$\mu = 1{,}57 \cdot 10^{-3}$,

$B_x^2 = B^2 + B_{CMB}^2$,

$B$ - is magnetic field strength of source, in $10^{-5}$ Gauss,

$B_{CMB} = 0{,}32(1+z)^2$ – is equivalent magnetic field strength corresponding to intensity of microwave background, in $10^{-5}$ Gauss,

$$E_i = \left(\frac{\nu_i}{1{,}41 \cdot 10^{18} B_x}\right)^{1/2}.$$

We consider the initial spectral index of sources in our sample has value $\alpha_0 = 0{,}8$ and the present spectral index of these sources has value $\alpha = 1{,}2$ (in case of S steep spectrum) and $\alpha = 2$ (in case of $C^+$ steep spectrum). As a result, at the frequencies $\nu_1 = 10$ MHz and $\nu_2 = 80$ MHz we obtain from (1) the estimation of synchrotron decay times at transient injection $t_{STI} \sim 10^8$ years (at the typical magnetic field strength $B$ (Miroshnichenko, 2012) of examined sources $B \sim 10^{-5}$ Gauss). This estimate is near to value of characteristic age $t$ (Miroshnichenko, 2013) for considered sources. So, the low frequency steep spectra of radio sources correspond to long evolution of these objects.

At process of synchrotron losses by relativistic electrons at their propagation in jets the radio lobes are formed with linear size of order near Mpc. Radio emission of these giant structures has maximum intensity at low-frequency band, particularly, at frequencies of decametre band.

For continuing study of evolution features of giant radio-sources with steep spectrum we examine ratio of emission of their radio lobes and accretion disk ($lg(S_{25}/S_{opt})$) versus jet propagation velocity $V_j$. Also, we consider ratio of emission of radio lobes and accretion disk crown ($lg(S_{25}/S_X)$) versus $V_j$. The value $lg(S_{25}/S_{opt})$ characterizes the ratio of monochromatic luminosities of source at the decameter band ( at the frequency 25 MHz) and the optical band (V). Analogously, $lg(S_{25}/S_X)$ corresponds to ratio of monochromatic luminosities of source at the frequency 25 MHz and te X-ray band (1 keV).

The jet propagation velocity of source $V_j$ (Miroshnichenko, 2016a) is determined from our estimates of linear size of radio structure $R$ (Miroshnichenko, 2012) and characteristic age of source $t$ (Miroshnichenko, 2013) as

$$V_j = R/2t \quad . \qquad (2)$$

Note, that derived estimate of jet propagation velocity of examined giant sources $V_j$ are subrelativistic ($\sim 0{,}1c$) (Miroshnichenko, 2016a). We obtained above-mentioned relations $\left[\lg\left(\frac{S_{25}}{S_{opt}}\right)\right](V_j)$, $\left[\lg\left(\frac{S_{25}}{S_X}\right)\right](V_j)$ for galaxies and quasars with linear steep spectrum S and break steep spectrum $C^+$ (Figure 1, Figure 2).For galaxies of both spectral types ($G_S$ and $G_{C^+}$) the positive trend is visible in relation $\left[\lg\left(\frac{S_{25}}{S_{opt}}\right)\right](V_j)$, that is, for greater values $V_j$ the greater values of radio emission of radio lobes and accretion disk are observed (Figure 1). Besides, it follows from Figure 1, that the relation for each spectral type of galaxies (S and $C^+$) looks as separate branch.

Quasars in our sample ($Q_S$ and $Q_{C^+}$) also display the positive trend in relation $\left[\lg\left(\frac{S_{25}}{S_{opt}}\right)\right](V_j)$ (Figure 2). Appreciably, that quasars with linear steep spectra ($Q_S$) are located at the region of greater $V_j$, while the most of quasars with spectrum $C^+$ occupy the region of smaller jet velocities $V_j$ (Figure 2). The ratio of emission of radio lobes and accretion disk crown ($lg(S_{25}/S_X)$) for galaxies and quasars of both spectral types ($G_S$, $G_{C^+}$, $Q_S$, $Q_{C^+}$) versus $V_j$ indicate on positive trend localization of jets with linear steep spectrum ($G_S$, $Q_S$) in region of greater values $V_j$ (Figure 3, Figure 4).

For contribution of emission of radio lobes relatively of accretion disk versus the characteristic age $\left[\lg\left(\frac{S_{25}}{S_{opt}}\right)\right](t)$ shows clustering of objects ($G_S$, $G_{C^+}$), $Q_S$, $Q_{C^+}$) in two regions of ages (Figure 5, Figure 6). Galaxies and quasars with break steep spectrum ($G_{C^+}$, $Q_{C^+}$) have characteristic age $t$ by one order greater than it for galaxies and quasars with S steep spectrum ($G_S$, $Q_S$). Mean values of characteristic age (Miroshnichenko, 2013, Miroshnichenko, 2016a,b) for objects of our sample are: $\langle t_{G_S} \rangle = 5{,}22(\pm 0{,}36) \cdot 10^7$ years, $\langle t_{Q_S} \rangle = 5{,}80(\pm 0{,}34) \cdot 10^7$ years, $\langle t_{G_{C^+}} \rangle = 4{,}66(\pm 0{,}41) \cdot 10^8$ years, $\langle t_{Q_{C^+}} \rangle = 2{,}92(\pm 0{,}25) \cdot 10^8$ years.

Relation of ratio of emission of radio lobes and accretion disk crown versus characteristic age of sources for examined galaxies and quasars (Figure 7, Figure 8) has form similar to relation $\left[\lg\left(\frac{S_{25}}{S_{opt}}\right)\right](t)$, $\left[\lg\left(\frac{S_{25}}{S_X}\right)\right](t)$ (Figure 5, Figure 6).

Note, that the mean value of ratio of emission of radio lobes and accretion disk of source $\left\langle \lg\left(\frac{S_{25}}{S_{opt}}\right) \right\rangle$ has greater value for galaxies and quasars with linear steep spectrum S as compared to objects with spectrum $C^+$. Namely, these values are: $\left\langle \lg\left(\frac{S_{25}}{S_{opt}}\right)_{G_S} \right\rangle = 5{,}64(\pm 0{,}12)$,

$\left\langle \lg\left(\frac{S_{25}}{S_{opt}}\right)_{Q_S} \right\rangle = 5{,}33(\pm 0{,}06)$,

$\left\langle \lg\left(\frac{S_{25}}{S_{opt}}\right)_{G_{C^+}} \right\rangle = 5{,}15(\pm 0{,}12)$,

$\left\langle \lg\left(\frac{S_{25}}{S_{opt}}\right)_{Q_{C^+}} \right\rangle = 5{,}00(\pm 0{,}10)$. As well, the mean value of ratio of emission of radio lobes and accretion disk crown $\left\langle \lg\left(\frac{S_{25}}{S_X}\right) \right\rangle$ has greater value for considered objects with linear steep spectrum S as compared to these with break steep spectrum $C^+$:

$\left\langle \lg\left(\frac{S_{25}}{S_X}\right)_{G_S} \right\rangle = 9{,}03(\pm 0{,}23)$,

$$\left\langle \lg\left(\frac{S_{25}}{S_X}\right)_{Q_S} \right\rangle = 8{,}45(\pm 0{,}14),$$

$$\left\langle \lg\left(\frac{S_{25}}{S_X}\right)_{G_{C^+}} \right\rangle = 7{,}89(\pm 0{,}33),$$

$$\left\langle \lg\left(\frac{S_{25}}{S_X}\right)_{Q_{C^+}} \right\rangle = 7{,}78(\pm 0{,}17).$$

Thus, the greater contribution of emission of radio lobes is observed for more young sources (with linear steep spectrum S) than for more old sources (with break steep spectrum $C^+$). Alternative reason of difference of these parameters may be the increase of contribution of emission of accretion disk and accretion disk crown for sources with steep spectrum $C^+$. Probably, this is connected with activity recurrence of nuclei of giant sources.

### 3. Conclusion

The examined giant radio sources – galaxies and quasars with both types of steep spectra (S and $C^+$) from the UTR-2 catalogue reveal a number of evolution features:

1. Positive correlation is determined for the ratio of emission of radio lobes and accretion disk versus jet propagation velocity of sources. Analogous correlation is observed for the ratio of emission of radio lobes and accretion disk crown versus jet propagation velocity. At that, the greater jet propagation velocities are typical for galaxies and quasars with steep linear spectra S.

2. Values of ratio of emission of radio lobes and accretion disk of examined sources are clustered in two regions of sources ages. Sources with $C^+$ type of steep spectrum have characteristic age by one order greater than these with steep spectrum S.

3. Contribution of emission of radio lobes relatively accretion disk decreases at the increasing of age of steep spectrum sources (or the contribution of accretion disk or crown increases) testifying to their activity recurrence.

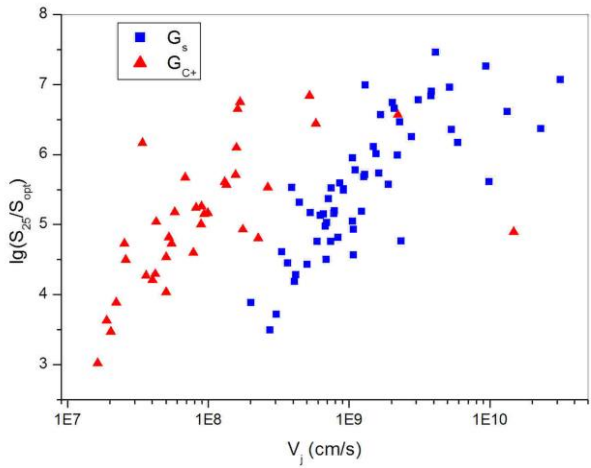

Figure 1: Ratio of emission of radio lobes and accretion disk versus jet velocity for galaxies with steep radio spectra of both types (S and $C^+$)

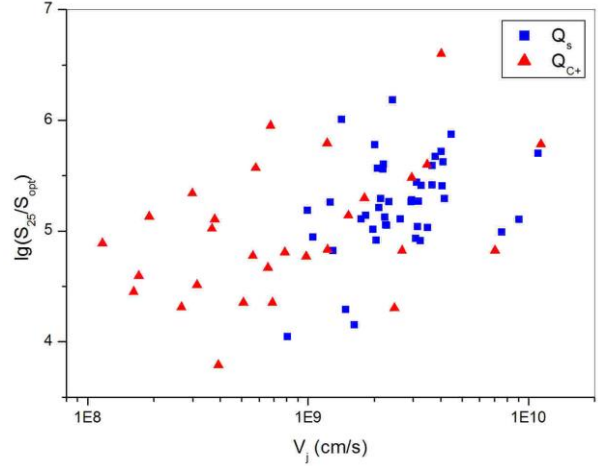

Figure 2: Ratio of emission of radio lobes and accretion disk versus jet velocity for quasars with steep radio spectra of both types (S and $C^+$)

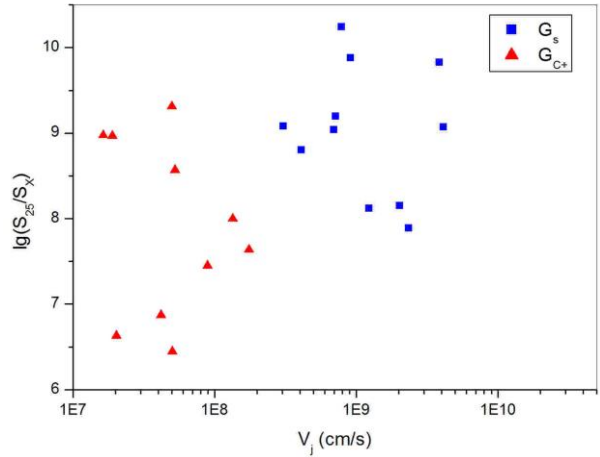

Figure 3: Ratio of emission of radio lobes and accretion disk crown versus jet velocity for galaxies with steep radio spectra of both types (S and $C^+$)

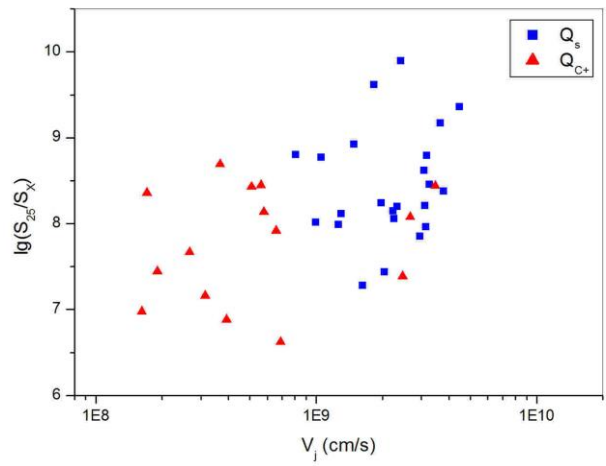

Figure 4: Ratio of emission of radio lobes and accretion disk crown versus jet velocity for quasars with steep radio spectra of both types (S and $C^+$)

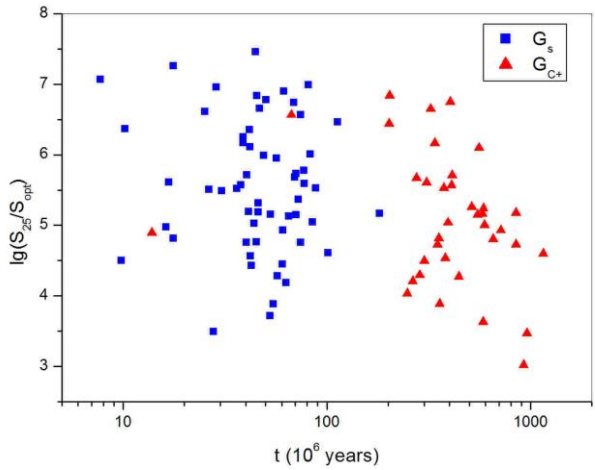

Figure 5: Ratio of emission of radio lobes and accretion disk versus characteristic age for galaxies with steep radio spectra of both types (S and C$^+$)

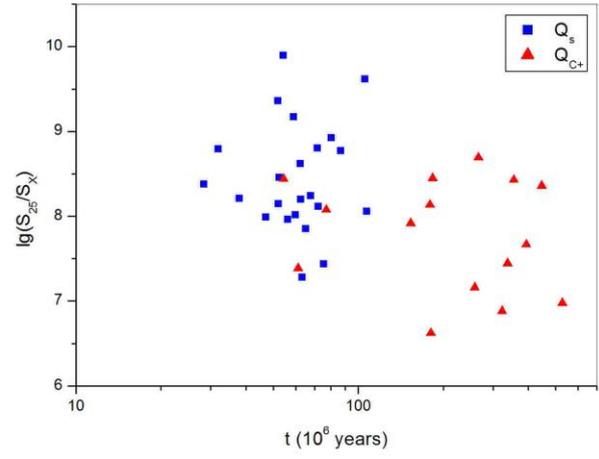

Figure 8: Ratio of emission of radio lobes and accretion disk crown versus characteristic age for quasars with steep radio spectra of both types (S and C$^+$)

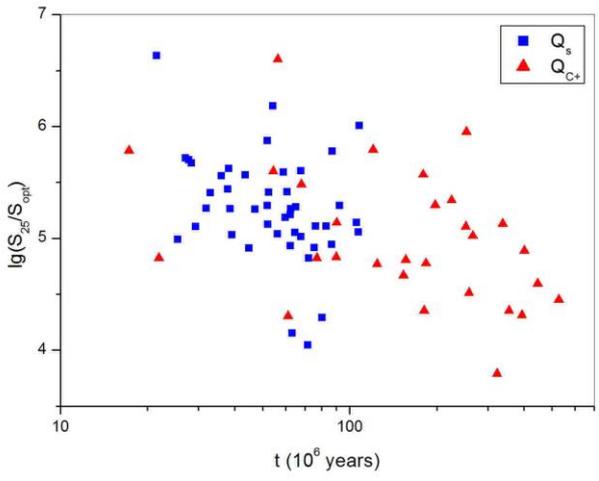

Figure 6: Ratio of emission of radio lobes and accretion disk versus characteristic age for quasars with steep radio spectra of both types (S and C$^+$)

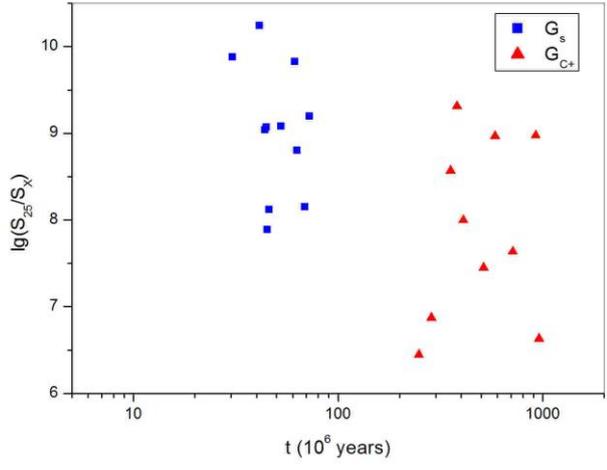

Figure 7: Ratio of emission of radio lobes and accretion disk crown versus characteristic age for galaxies with steep radio spectra of both types (S and C$^+$)

### References

Braude S.Ya, Miroshnichenko A.P., Sokolov K.P. & Sharykin N.K.: 1981, *Astrophys.Space Sci.*, **76**, 279.

Braude S.Ya., Miroshnichenko A.P., Rashkovski S.L. et al.: 2003, *Kinem.&Phys.Celest.Bodies*, **19**, 291.

Gorbatsky V.G.: 1986, Introduction to Physics of Galaxies and Clusters of Galaxies. (Nauka, Moscow).

Kardashev N.S.: 1962, *Sv.Astron.*, **6**, 317.

Miroshnichenko A.P.: 2012, *Radio Phys.&Radio Astron.*, **3**, 215.

Miroshnichenko A.P.: 2013, *Odessa Astron.Publ.*, **26/2**, 248.

Miroshnichenko A.: 2014, Independent Estimates of Energies of Steep-Spectrum Radio Sources, in Multiwavelength AGN Surveys and Studies, eds. A. Mickaelian & D. Sanders, (Cambridge University Press, Cambridge), 96.

Miroshnichenko A.P.: 2016a, Abstracts of VIII Conference "Selected Issues of Astronomy and Astrophysics", I.Franko National University of Lviv, 20.

Miroshnichenko A.P.: 2016b, *Odessa Astron. Publ.*, **29/2**, 173.